\documentclass[12pt,a4]{revtex4}
\usepackage{graphicx,subfigure}
\usepackage{dcolumn}
\usepackage{graphicx}
\begin{document}
\title{\Large \bf Sympatric Speciation in a Simple Food Web}
\author{{\bf K. Luz-Burgoa$^{\dag}$, Tony Dell$^{**}$ and Toshinori Okuyama$^{\ddag}$}
\email{karenluz@if.uff.br, tony@alguna.cosa, toshi@otra.cosa}\\
{\small $\dag$ Universidade Federal Fluminense, RJ, Brazil, karenluz@if.uff.br}\\
{\small $**$ Department of Tropical Biology, James Cook University, Townsville, QLD 4811, 
Australia}\\
{\small $\ddag$ University of Florida, Zoology Department. University of Florida
Gainesville, FL}\\
{\small Contribution to the Proceedings of the Complex Systems Summer School 2004,}\\ 
{\small organized by the Santa Fe Institute, New Mexico USA, August 24, 2004}
}
%\date{}

\maketitle
{\bf Abstract: Observations of the evolution of species groups in nature, such as well 
recognized Galapagos finches, have motivated much theoretical research aimed at understanding 
the processes associated with such radiations. The Penna model is one such model and has 
been widely used to study aging. In this paper we use the basic Penna model to investigate 
the process of sympatric speciation in a simple food web model. Initially our web consists  
of a primary food source and a single herbivore species that feeds on this resource.  
Subsequently we introduce a predator that feeds on the herbivore. In both instances we 
directly manipulate the food source, that is, its size distribution, and monitore the  
changes in the populations structures. Sympatric speciation is obtained for   
the consumer species in both webs, and our results confirm that the speciation velocity  
depends on how far up,in the food chain, the focus population is feeding. Simulations 
are done with three different sexual imprinting-like mechanisms, in order to discuss adaptation 
by natural selection.}

\bigskip

{\center{\Large\bf Introduction}}\\
According to classical theories of speciation, mating signals diversify, in part, as an 
incidental byproduct of adaptation by natural selection to divergent ecologies~\cite{darwin}. 
Nevertheless 
empirical evidence in support of this hypothesis is limited~\cite{rhe47,crrjn411,acs423} 
and is a matter of controversy\cite{cfevecre5}.\\

{\it \large{A. Observations and Measurements in Nature}}\\
A great living example of evolution in action are Darwin's Finches, a group of 13 finch 
species of the Galapagos Islands~\cite{bgs214}. The beaks of each species is 
apparently specifically adapted to feed on a precise food type, running from seeds  
and cactus flowers to buds or insects. It is supposed that natural selection drove, 
and is indeed still driving, the beak morphology of each finch species. Detailed analysis 
of these birds have revealed that the changes in beak morphology can occur very quickly, 
even within the course of a single season~\cite {sgs227}.

The sexual imprinting-like mechanism is 
apparently ubiquitous in Darwin's finches and is present in some form species of all orders 
of birds examined so far~\cite{ggs256}. Song is a culturally transmitted trait learned 
during a short sensitive period early in finche's life, and later used in courtship and 
mate choice. Males typically sing a simple, short song, and retain it unaltered throughout 
life. Females do not sing but do learn songs that are later used in mate choice of 
hybridizing birds and of the hybrids themselves~\cite{ggbjls76,gge50}. It has been shown 
that as a consequence of beak evolution there have been changes in the structure of finch 
vocal signals (Podos~\cite{pn409}).

The diversification of beak morphology and body 
size of the finches has shaped patterns of vocal signal evolution, such birds with large 
beaks and body sizes have evolved songs with comparatively low rates of syllable repetition 
and narrow frequency bandwidths. Patterns of correlated evolution among morphology and 
song are consistent with the hypothesis that beak morphology constrains vocal evolution. 
Different beak morphologies differentially limit a bird's ability to modulate vocal tract 
configurations during song production. Data~\cite{pnbio54,psr207} illustrate how 
morphological adaptation may drive signal evolution and reproductive isolation, and 
furthermore identify a possible cause for rapid speciation in Darwin's finches. \\

{\it{\large B. Theory}}\\
Traditionally, two main classes of models are used to explain speciation. Allopatric 
speciation models assume that the initial population is suddenly divided into two 
geographically isolated subpopulations, which then diverge genetically until they 
become reproductively isolated. However many migratory birds do not seem to fit the 
basic requirement of long periods of geographical isolation needed for allopatric 
speciation, which led to the proposal of a sympatric speciation mechanism. Sympatric 
speciation corresponds to the division of a single local population into two or more 
species. Understanding how sympatric 
speciation can be driven has thus attracted much theorical effort\cite{cfevecre5}.

According to Darwin~\cite{darwin}, sympatric speciation is driven by disruptive, 
frequency-dependent 
natural selection caused by competition for diverse resources. Recently several authors 
have argued that disruptive sexual selection can also cause sympatric speciation. Here, 
we use the Penna model to examine this process~\cite{penna}. The model assumes that 
competition for resource and sexual selection are the dominant forces acting on the 
population. We explore sympatric speciation within simple food webs with different 
sexual imprinting-like mechanisms.  
\bigskip

{\center{\Large\bf Model description}}\\
The Penna model for biological aging is based entirely on Darwinian evolution and mutations. 
Originally focused on problems of biological aging, applications to several different 
evolutionary problems substantially increased its scope~\cite{sjpkatcsbook}.\\

{\it{\large A. Penna Model}}\\
In the sexual version of the Penna model used here, each individual has two bitstrings 
inherited from mother and father, respectively. Gametes 
(single bitstrings) are produced by random crossover between these two bitstrings,
followed by one random mutation. Each female of age ten or above tries
many times to find randomly a male aged ten or above for mating, and if she
succeeds she gets two offspring, having one of the father's gametes and one of the
mother's gametes as its two bitstrings.
The offspring's sex is fixed randomly. If at a specific bit position, one of the two 
bitstrings has a bit zero and the other has a bit one, it 
affects the health of that individual if and only if this position is one for 
which the harmful allele (bit 1) is dominant. Ten out of the 32 
possible positions are randomly selected as dominant, the remaining 
22 as recessive. There is a competition for space and food given by the logistic Verhulst 
factor. The complete Fortran program is listed in \cite{book}.\\

{\it{\large B. Speciation Model}}\\
In the first simulations using the Penna model with phenotype to get sympatric 
speciation 
~\cite{medeiros,ksjsabjp33} it has been considered that competition for 
resources changes according to the ecology. 
In our model the competition does not change, and fitness and mate choice depend on the same 
trait. A new pair of non-age structured bitstrings is added to the original one, to represent 
the individual's phenotype. The phenotypic characteristic is measured by counting 
the number of recessive bit positions (choosen as 16), where both bits are set to 1, plus 
the number of dominant positions with at least one of the two bits set. It will therefore 
be a number $k$ between 0 and 32. The mutation probability per locus of this phenotype is 
set to $0.5$ in all simulations.\\
The death probability by intraspecific competition, for extremal phenotypes, is given by:
\begin{equation}\label{fontes} 
V_{<(>)}(t) = \frac{pop_{<(>)}+pop_m}{\left(  Capacity*SourceD(k)\right)}
\end{equation}
where $pop_{<(>)}(t)$ accounts for the population with phenotype $k$ $<16$ and $k>16$, 
respectively. The Verhulst factor for intermediate ($m$) phenotypes is:
\begin{equation}\label{fontes} 
V_m(t) = \frac{pop_m+(pop_<+pop_>)*0.5}{\left( Capacity*SourceD(k)\right)}
\end{equation}

$Capacity*SourceD(k)$ is the carrying capacity of the environment as seen by each individual, 
since it depends on the 
number $k$. At every time step, and for each individual, a random number is generated;  
if this number is smaller than $V$, the individual dies.  
In both cases presented in the next section, $SourceD(k)$ is the first species of 
a chain food. It may, for instance, represent plants with a given size distribution. 
Individuals with extremal phenotypes ($pop_<$, $pop_>$) compete for small/large plants 
among the individuals with its same extremal phenotype,   
and also with the whole intermediate population (eq. 1). Individuals with intermediate 
phenotypes ($pop_m$) compete among themselves and also with half of each population  
presenting an extremal phenotype (eq. 2).
\bigskip

Finally we refer to mating selectiveness, where we introduce into each genome a
locus that codes for this selectiveness, also obeying the general rules of the 
Penna model for genetic heritage and mutation. If it is set to $0$, the individual
is not selective in mating (panmictic mating). It is selective
(assortative mating) if this locus is set to $1$. At the beginning of the 
simulations all females are non-selective. The mutation probability  
for this locus is set to $0.001$. Females that are selective
choose mating partners according to one of the following mating strategies.\\
{\it{ Mating strategy 1}}\\
If a female has phenotype $k<16$($\geq 16$) it prefers a male with phenotype 
$k<16$($\geq 16$). This sexual imprinting-like mechanism is ubiquitous in female.\\
{\it{ Mating strategy 2}}\\
In this case a female chooses, among six males, the one with the smallest difference between 
its phenotype $k_F$ and the male's phenotype $k_M$.\\
{\it{ Mate strategy 3}}\\
The mating of a pair occurs with probability $=(k_F-k_M)/32$, where $k_F$ is the 
female phenotype and $k_M$ is the male phenotype.\\

{\center{\Large\bf Results and Discussion}}\\

\begin{widetext}
\begin{figure}
\begin{center}
\subfigure{
 \includegraphics[width=6.2cm,angle=270]{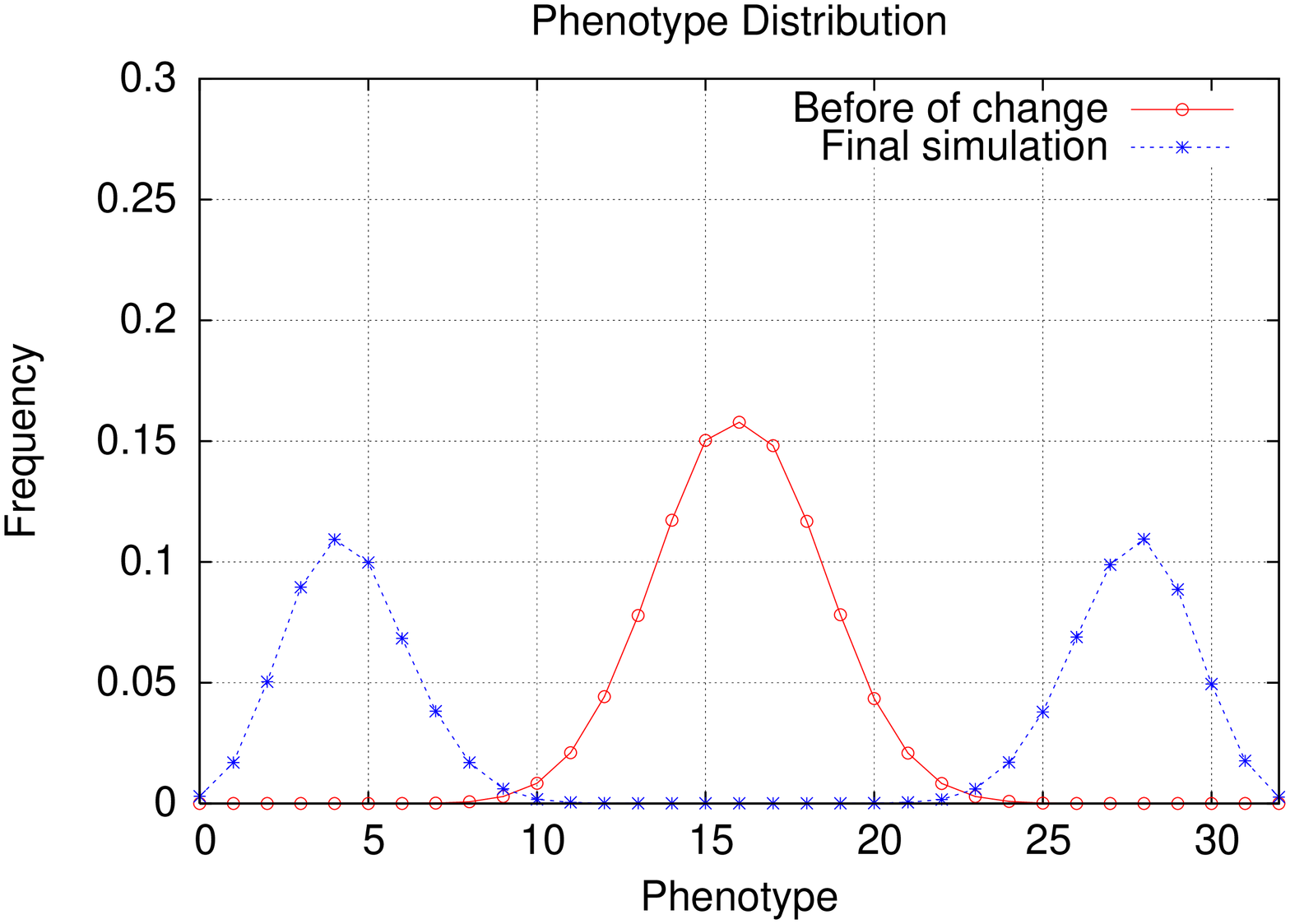}}\subfigure{
 \includegraphics[width=6.2cm,angle=270]{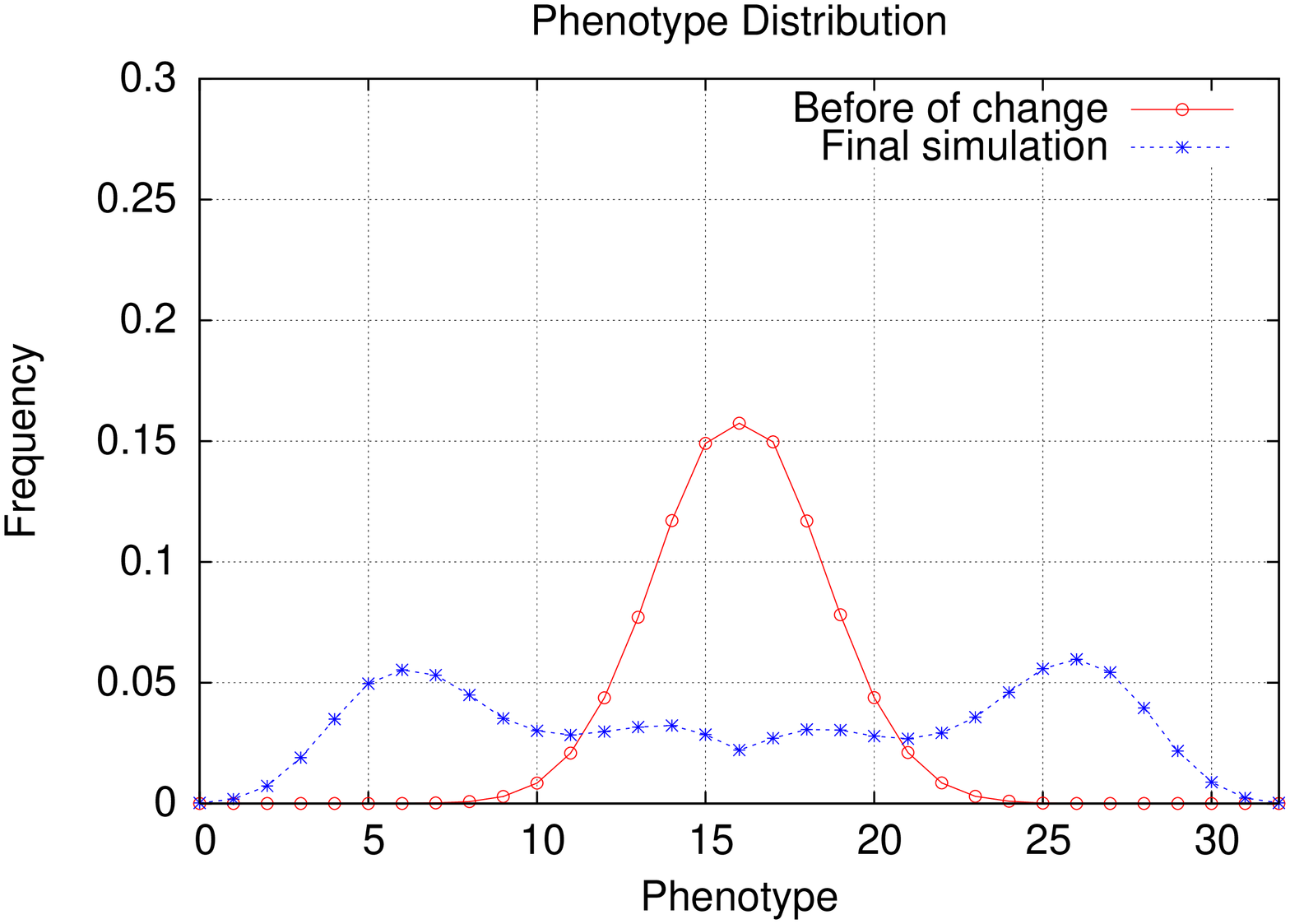}}
\end{center}
\caption{Phenotypic distributions of the second species, in a food chain of two species, for  
mating strategy 1 or 2 (left part) and for mating strategy 3 (right part), before and 
after the sudden change of the first species distribution.}
\label{fig:dois}
\end{figure}
\end{widetext}

{\it{\large A. Two species food web}}\\
Here the consumer (a herbivore) has genetical properties and evolves 
for 250 generations with a constant food distribution, which is the first species. Suppose 
for instance, that during a given season this food distribution consists of plants which   
sizes favor the individuals of the second species presenting medium phenotypes.
Suddenly, due to a new different rainfall regime, the first species distribution 
changes into a bimodal one, now favoring individuals with extremal phenotypes: 
\begin{equation}\label{plantas}
SourceD(k)~=\left\lbrace\begin{array}{ll} ~1.0~-~\frac{|16~-~k|}{20.0}&Before\\ 
~0.1~+~\frac{|16~-~k|}{20.0} & After \end{array}\right. 
\end{equation}
For the mating strategies 1 and 2 the second species phenotypic distributions are the same,   
left part of fig.\ref{fig:dois}. This is an interesting result since in strategy 1 
the female knows the 
drift direction of the ecology and it is easy to understand why the population presents two 
substantially different phenotypes and how reproductive isolation between them has 
driven the elimination of all intermediate phenotypes. Females with mating strategy 2 do not 
know this direction and, even so, the ecology is driving their preferences in the same way 
as with choice 1.
%fig1

For the mating strategy 3 (rigth part of fig.\ref{fig:dois}) there is no correlation between 
ecological 
changes and female preferences, and the intermediate phenotypes are not totaly eliminated.   
However, this strategy is more realistic than strategy 2, since the female's preference  
is subject to the males availability.\\

\begin{figure}[ht]
\includegraphics[width=5.8cm,angle=270]{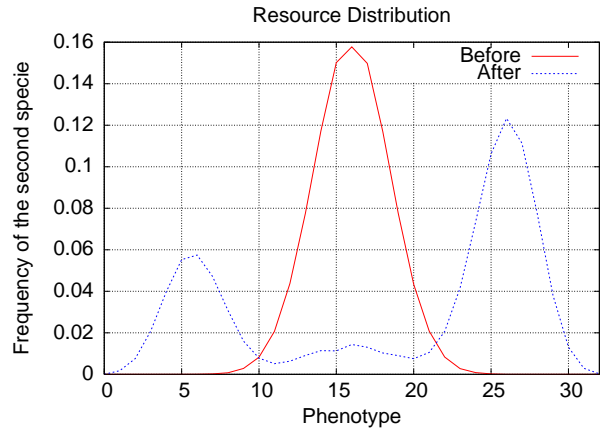}
\caption{Phenotypic distribution of the second species in a food chain of three 
species; there is no sexual selection for this species.}
\label{di}
\end{figure}
%fig2

{\it{\large A. Three species food web}}\\
The consumer (a predator) feeds solely on the hervibore and has genetical properties. 
The hervibore also has genetical properties, but no mating preference, and evolves for 
250 generations with a given food distribution - the first species (which consists of 
the same plants of the previous case). Now, when the first species distribution suddenly 
changes, the phenotypic distribution of the second 
species, as a consequence, also changes - fig.\ref{di}. In this figure  
the red distribution is stable but the blue one is not and sometimes there are  
more individuals with one of the extremal phenotypes than the other.

\begin{widetext}
\begin{figure}
\begin{center}
\subfigure{
 \includegraphics[width=6.2cm,angle=270]{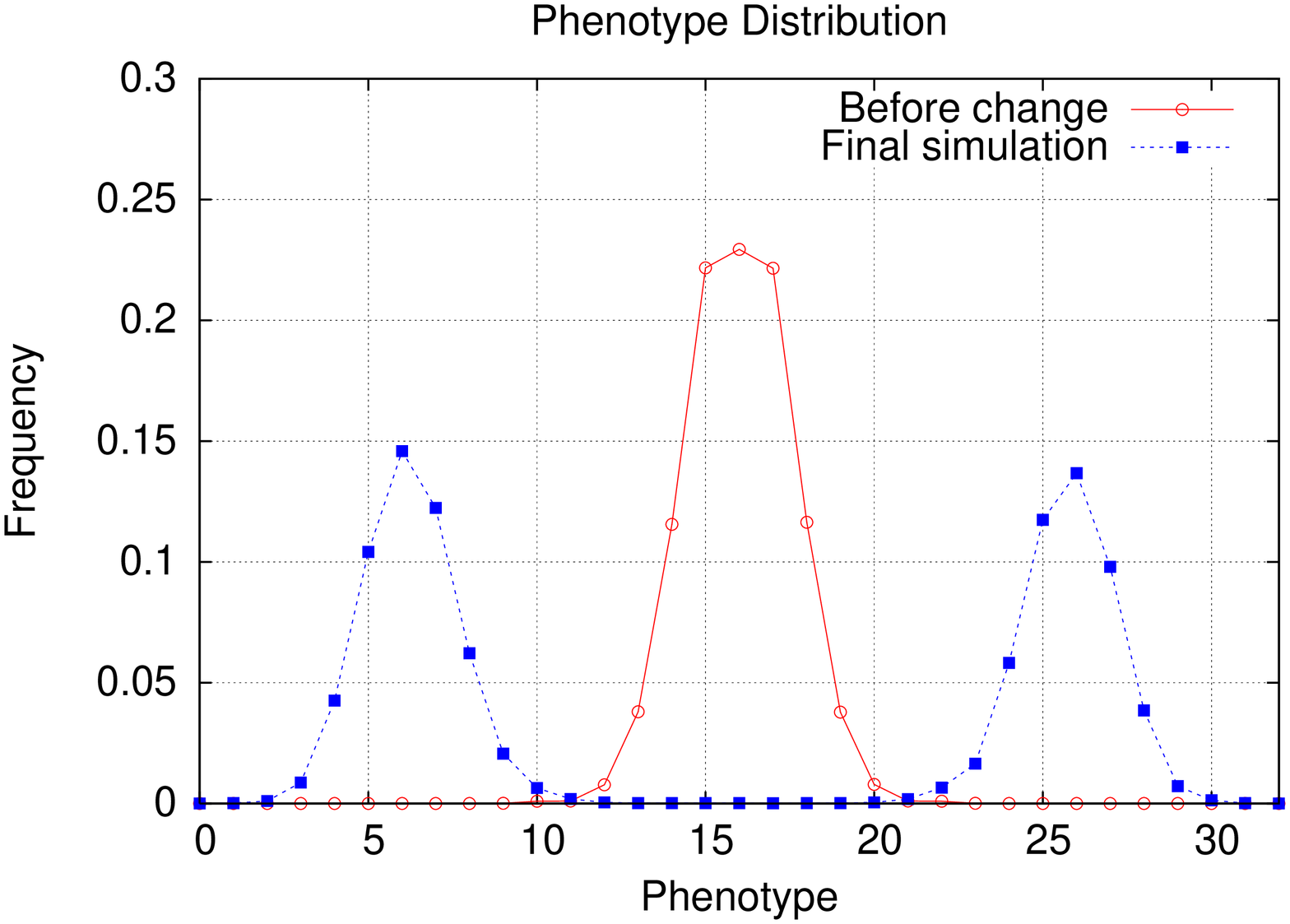}}\subfigure{
 \includegraphics[width=6.2cm,angle=270]{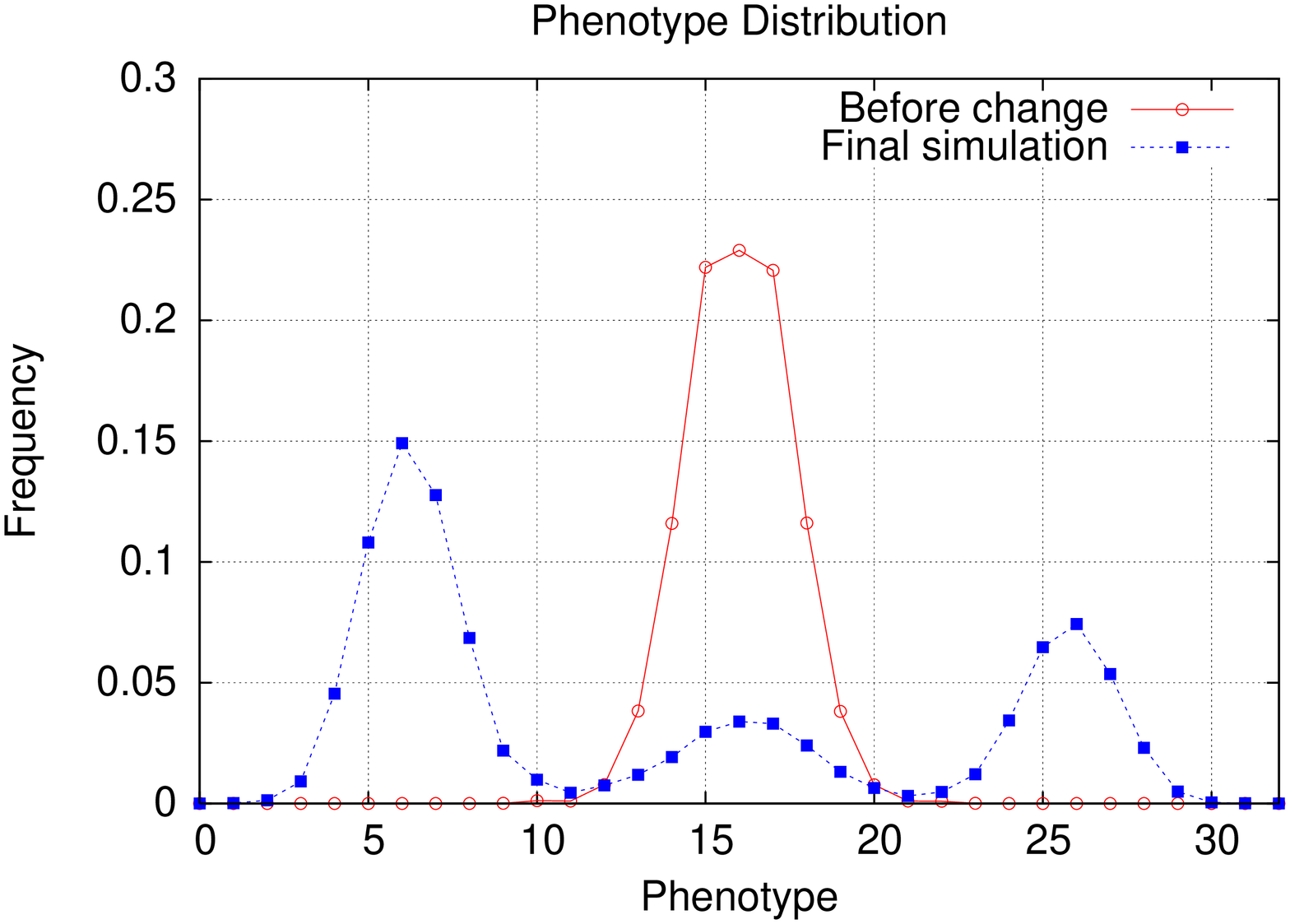}}
\end{center}
\caption{Third species phenotypic distributions for mating strategies  
1 or 2 (left part) and mating strategy 3 (right part).}
\label{fig:3}
\end{figure}
\end{widetext}

The effects of the different mating strategies for this food chain are the same 
as those of the previous one - fig.\ref{fig:3}. 

%fig3

\begin{figure}[ht]
\includegraphics[width=5.8cm,angle=270]{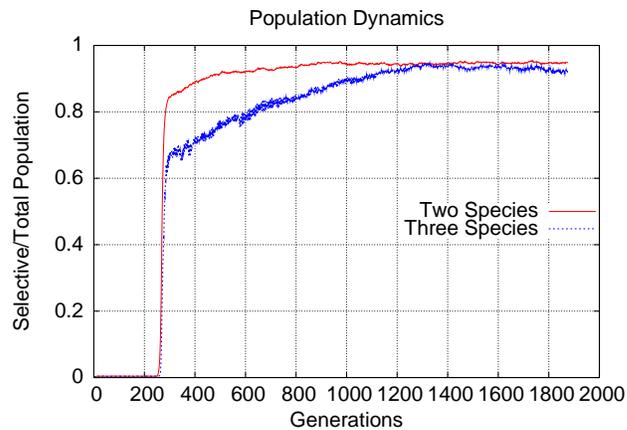}
\caption{Comparison between fractions of selective individuals using mating strategy 1, 
for the two different food chains.}
\label{co}
\end{figure}

The most important difference between the two food chains is the speciation velocity,  
measured through the time evolution of the fraction of selective individuals in the 
populations. Fig.\ref{co} shows that  
intermediate phenotypes desappear faster in the two species food chain than in the 
three species one.

%fig4

Another difference is in the variance of the distributions per phenotype of the two chains  
(left parts of figs.~\ref{fig:dois} and \ref{fig:3}). The red and blue distributions for the 
two species chain present a larger variance than the equivalent ones for the three 
species chain. Also the mean phenotypes  
after speciation (blue distributions) in the two species chain are at 4 and 28, while in 
the three species case, at 6 and 26.
It means that the mean difference between extremal phenotypes for the two species chain 
is smaller than for three species one.\\ 
These results are limited  to be compared with those from 
real finches observations, not only  
because we are using a toy model but mainly because natural evolution is a too complicated 
process!!.

{\center{\Large\bf Acknowledgements}}\\
We thank the Santa Fe School 2004 and all their participants, S. Moss de Oliveira for fruitful discussions, 
Constantino Tsallis's help and the agencies Conselho 
Nacional de Desenvolvimento Cientifico e Tecnologico(CNPq) and Centro Latino Americano 
de F\'{\i}sica (CLAF) for financial support.\\

\end{document}